\documentclass{PoS}
\usepackage{graphics}
\usepackage{graphicx}
\usepackage{caption}
\usepackage{subcaption}
\title{Confinement in large N gauge theories}

\ShortTitle{Confinement in large N gauge theories}

\author{\speaker{Antonio Gonz\'alez-Arroyo}$^{ab}$ and Masanori
Okawa$^c$\\
\llap{$^a$}Instituto de F\'{\i}sica Te\'orica UAM/CSIC\\
\llap{$^b$}Departamento de F\'{\i}sica Te\'orica, C-15 \\
       Universidad Aut\'onoma de Madrid, E-28049--Madrid, Spain \\
       \llap{$^c$}Graduate School of Science, Hiroshima University\\
       Higashi-Hiroshima, Hiroshima 739-8526, Japan\\
       E-mail: \email{antonio.gonzalez-arroyo@uam.es},
       \email{okawa@sci.hiroshima-u.ac.jp}}

\abstract{We report some recent results obtained for large N gauge
theories which support the idea of volume reduction. Results for the
string tension of the Twisted Eguchi-Kawai model match with those
obtained from extrapolation from finite N. Determination  of other
observables is currently under way. Application of the twisted
reduction idea to the case of 1 or 2 flavours of quarks in the adjoint
representation of the group offers promising results. Preliminary
results for the string tension point towards a very different behaviour
for 1 and 2 flavours. While the string tension remains finite for 1
flavour at the critical massless quark limit, it seems to vanish with
a large anomalous dimension for the $N_f=2$ case. This is consistent with 
the predicted Infrared fixed point expected in the latter case.}

\FullConference{Xth Quark Confinement and the Hadron Spectrum,\\
		October 8-12, 2012\\
		TUM Campus Garching, Munich, Germany}

\begin{document}

\section{Introduction}
The idea of playing with the rank of the symmetry group as a parameter
that can help to better understand  the behaviour of a dynamical system, 
first appeared in the context of statistical mechanics. A large value
for this rank suppresses fluctuations and the mean field approximation
becomes increasingly accurate. `t Hooft\cite{thooft} applied the same idea to the gauge
group of a gauge invariant theory, but this time not all fluctuations
are suppressed and the theory remains non-trivial. To date, the large N
limit of SU(N) Yang-Mills theory remains unsolved. Nevertheless, there
are several exact results that have been obtained (or conjectured) for 
this limit, which amount to a considerable simplification of the
dynamics. For example, in perturbation theory only a subset of Feynman
diagrams have to be considered at leading order in 1/N. Furthermore, 
quark loops for a finite number of flavours are suppressed, and the
quenched approximation is correct. This is only true if quarks are in
the fundamental representation as for QCD. Fermionic fields in two-index 
representations~(symmetric, antisymmetric or adjoint) are not suppressed, 
so that the large N dynamics is different in each case. There are also
a large number of non-perturbative results that have been established
or proposed. 

Despite the time that has passed since the idea was first presented,
the interest in large N gauge theories has not decreased. There are
many hints that the dynamics in this limit is even quantitatively
close to that of low rank theories as QCD. This makes it interesting 
at the phenomenological level. At a more conceptual level, the
simplifications achieved at large N seem strong enough to hint that
the theories might be, at least partially, solvable. These
simplifications are not achieved at the cost of changing the basic
dynamical properties of the theory: Confinement or chiral symmetry
breaking etc are not lost at large N. 

There is also a fascinating connection among ordinary quantum field
theories and string theories which goes under the name {\em AdS/CFT}\cite{Maldacena}.
Again the large N limit implies an important simplification, leading to
a classical description of the string side. As a matter of fact, the connection
between the large N expansion and string theory is already suggested 
at the perturbative level: the 1/N classification of Feynman
diagrams follows the structure of the topological expansion of
two-dimensional surfaces (the world sheet of a string). 

Certain properties of large N theories adopt the form of
conjectures, which rely on non-perturbative properties of the vacuum. 
Some of these are fairly recent, like the orientifold planar
equivalence~\cite{OPE}, which relates the dynamics of quarks in different
types of two-index representations. 

Testing these conjectures and extracting the spectra and other
non-perturbative properties of large N gauge theories is clearly an
important goal in quantum field theory. Unfortunately, the most
powerful model-independent methodology ---lattice gauge theories---
does not seem to inherit the simplificatory character of the large
colour limit. Indeed, the standard road to study these theories is by
extrapolating the results of finite N gauge theories. A priori, the
larger $N$ the more computationally demanding is its study.  Despite
this fact, there have been many interesting studies indicating a fast
approach to large N. For a recent review one can consult
Ref.~\cite{panero}.

One of the earlier conjectures, known as {\em reduction}, seems to
change this general pattern. The idea, put forward in
Ref.~\cite{EK}, is that at large N results become volume
independent. The derivation is done within the lattice approach and
under the assumption that the $Z(N)^d$ symmetry of the theory is
respected ($d$ is the space-time dimension). Reduction  has important 
conceptual and practical consequences, since the computational cost 
of higher $N$ can be compensated by simulations done in smaller boxes.
The extreme limit is that of a one-point lattice, and we end up with 
a matrix  model of $d$ matrices. The original proposal of
Ref.~\cite{EK}, known as Eguchi-Kawai model, was soon
disproved~\cite{BHN}, since the assumption of $Z(N)$ symmetry fails at
weak coupling and $d>2$, where the continuum limit is obtained. The authors 
of Ref.~\cite{BHN} proposed a modification called Quenched
Eguchi-Kawai model which seemed to circumvent this problem, but has
recently been shown to fail too. 

\section{The Twisted Eguchi-Kawai model}

The present authors proposed a different modification called the Twisted
Eguchi-Kawai model (TEK)\cite{TEK}, which also seemed to avoid the symmetry
breaking problem. The idea is simple: formulate the finite volume
lattice theory using twisted boundary conditions, and then reduce the
model to a single point. The partition function is given by:
\begin{equation}
Z=\prod_{\mu} \int dU_\mu \, \exp\{-S_{\mathrm{TEK}}(U_\mu,Z_{\mu \nu})\}
\end{equation}
where the integral is over $d$ $N\times N$ unitary matrices and the
action is given by 
\begin{equation}
S_{\mathrm{TEK}}(U_\mu,Z_{\mu \nu})=-b \sum_{\mu \nu} Z_{\mu \nu}
\mathrm{Tr}(U_\mu U_\nu U_\mu^\dagger U_\nu^\dagger)  
\end{equation}
The constants $Z_{\mu \nu}$ are elements of the center and can be
parameterised as $\exp\{2 \pi i n_{\mu \nu}/N\}$ in terms of an integer
valued antisymmetric tensor $n_{\mu \nu}$. Different values for this
tensor correspond to different boundary conditions. The particular
choice $n_{\mu \nu}=0$ ($Z_{\mu \nu}=1$) corresponds to the original  
Eguchi-Kawai model. The weak coupling behaviour depends crucially on
the value of $n_{\mu \nu}$. In four space-time dimensions, a particularly 
symmetric choice is achieved by taking $N=L^2$ and $n_{\mu \nu}=kL$ for
$\mu<\nu$.  In this case, at weak-coupling the $Z(N)^4$ symmetry group 
is broken down to $Z(L)^4$, which is enough to preserve the reduction
idea as $L\longrightarrow \infty$ ($N\longrightarrow \infty$).

The perturbative expansion of the TEK model gives rise to a Lie
algebra expansion which mimicks  the momentum  expansion of a $L^4$ 
lattice gauge theory. The propagator just coincides with that of an $L^4$  
volume, and  the vertices contain momentum dependent phases which lead
to rapidly oscillating factors for non-planar graphs. Indeed, the
continuum version of the TEK model~\cite{AGAKA} produces  a realization 
of non-commutative field theory~\cite{nekrasov}. This provides an 
explanation of the reduction idea at the perturbative level, which
complements the Wilson loop equation of motion derivation of Eguchi
and Kawai.

However, in the last few years there have been several results which 
challenge the validity of the TEK model. On one side, several
numerical results~\cite{IO}-\cite{TV}-\cite{AHHI} showed, that for
large enough values of L,  $Z(L)^4$ symmetry breaks down at intermediate
values of the coupling
$b$ for the symmetric twist with minimal flux $k=1$. On the other hand, 
instabilities in the perturbative expansion of non-commutative field 
theories were reported~\cite{GHLL}, which might transmit to instabilities in the
$Z(L)$-symmetric vacuum  of the TEK model~\cite{BNSV}. 

In a previous paper~\cite{AGAMO}, the present authors analyzed the problem and
proposed that, in order to avoid symmetry breaking, the large N limit
has to be taken by scaling $k$ with $L$ rather than keeping $k$  fixed. 
In addition, to avoid instability problems of the type observed in 
non-commutative studies one should also keep $\bar{k}/L$ finite, where 
$\bar{k}$ is defined by $k \bar{k} =1 \bmod L$. With these
prescriptions incorporated, all  numerical studies done at even higher
values of $L$ have shown no sign of symmetry breaking or instability.

The issue is far from being settled at the fundamental level. A full
understanding of whether and how reduction takes place for the pure
gauge theory is still missing. For practical reasons one should also
understand the size and nature of the 1/N corrections compared to the 
finite volume counterparts. A systematic study in this direction is 
hard, so that the authors decided to attack the simpler case of 
2+1 dimensions first. A preliminary account of our findings has already
appeared~\cite{MGPAGAMO} and a full paper on the subject will soon
follow. In that work we consider different fluxes $k$, different values 
of $N$ and of the spatial size $l$. Perturbative calculations show that 
self-energy problems reported for non-commutative field theories are
absent when the flux is scaled as proposed in Ref.~\cite{AGAMO}.

\section{New results for Pure gauge theory at large N}
From a more practical standpoint it is much better to put the ideas to
work in computing physical quantities of large N gauge theories with
or without the use of reduction. For that purpose we set up a program 
to determine the string tension of the large N continuum Yang-Mills
theory. Very nice previous analysis existed, of which we 
will single out that of Ref.~\cite{Allton} which used  Polyakov loop
correlation functions. For technical reasons we decided to use ordinary 
Ape-smeared Wilson loop expectation values. The goal was two-fold. On
one hand we tried to develop a  technique to extract precise values of 
the string  tension from Creutz ratios of smeared Wilson loops. Once,
these values were obtained  for various $N$, the large N string tension 
value would be obtained by extrapolation. On the other hand, we would
use the TEK model with our prescription of fluxes $k$ to obtain what
would tentatively be an alternative determination of the large N
string tension. 

The main results  of the previous analysis have been recently 
presented\cite{AGAMOII}. Results for groups SU(3), SU(4), SU(5), SU(6)
and SU(8) fall nicely in the curve
$\frac{\Lambda_{\bar{\mathrm{MS}}}}{\sqrt{\sigma}}= 
0.515(3)+0.34(1)/N^2$. On the other hand the TEK model with N=841 has 
$\frac{\Lambda_{\bar{\mathrm{MS}}}}{\sqrt{\sigma}}=0.513(6)$. Systematic
errors entering in the definition of the scale are not included in the
errors given, but do not affect the relative difference between the
extrapolated and TEK result. Alternatively, one might determine the
scale from the data itself, using a nonperturbative renormalization
prescription. Once more, the conclusion is that the string tension for
the TEK model matches the extrapolated one within the 2\% error.  This
provides a numerical support for the validity of the reduction idea,
implemented through the twisted one-site model.  

Our work contains also interesting results concerning corrections to
the area law behaviour of Wilson loops. This has implications for 
the effective string theory description of Wilson loops. It seems that 
the aspect ratio dependence of Wilson loops cannot be described entirely
in terms of string fluctuations. One gluon exchange contributions turn
out to be quite relevant. This is not surprising since a similar
effective description including both singlet stringy components plus
perturbative corrections has been found to be phenomenologically
relevant since early times. Furthermore, there is a recent
paper~\cite{Lohmayer} in which, using a quite different but complementary 
analysis of Wilson loops at large N, the authors present results and
conclusions in agreement with ours.

Up to now, the only physical quantities that have been analyzed with the use 
of the reduced model are the string tension and the asymptotic parameters
mentioned before. However, a priori, the TEK model can be used to
determine several other interesting quantities. We are currently
exploring the determination of the chiral condensate, the finite
temperature phase transition and the meson spectrum. At first, one
might be puzzled by the possibility of determining these quantities
with a single point matrix model. As an example, let us illustrate for 
example a possible  program to determine the deconfinement temperature
$T_c$. 

The standard way to determine the deconfinement temperature is to
study an asymmetric box  $L_0\times L_s^3$ for $L_0\ll L_s$. As the
coupling $b$  crosses  a certain critical value $b_c$, the expectation value
of time-like Polyakov loops ceases to vanish. The critical temperature
is then given by $T_c=/(L_0 a(b_c))$, where $a(b)$ is the lattice
spacing in physical units. How can one obtain such a thing in a 1-point box?
Indeed, there are two ways. One way is to modify the twist to make it
asymmetric~\cite{KvB}. The other way, which we have explored, is to
preserve the symmetric twist, but  make the spatial $b_s$ and time-like
$b_0$ couplings different. This generates also different scales 
$a_s(b_0,b_s)$ and  $a_0(b_0,b_s)$. Then a transition must occur as 
$L a_0(b_0,b_s)=1/T_c$ (where $L=\sqrt{N}$). How would one notice the
transition? We recall that we need vanishing Polyakov loops for
reduction to operate. However, at weak coupling we know that the
symmetry group breaks down to $Z(L)^4$. In the asymmetric coupling 
situation the deconfinement transition is signalled by the breaking
from $Z(N)$ to $Z(L)$ in the temporal direction. The order parameter
is then given by $\mathrm{Tr}(P_0)/N$, where $P_0=U_0^L$ is the
effective temporal  Polyakov loop. 

Does this happen in the reduced model with asymmetric couplings? The
answer is yes, as exemplified in Fig.~1, describing the $b_0/b_s=4$ case. 
In it  we show the Monte Carlo history  of the modulus of $\mathrm{Tr}(U_\mu^L)/N$. For smaller 
coupling, all expectation values are consistent with zero (shown by a small value of
the modulus). This is given by the upper
curves which are displaced by 0.25 to make them visible. A tiny change
in $b=\sqrt{b_0 b_s}$   then
generates a situation displayed in the bottom curves. The spatial
observables remain small and the temporal one becomes non-zero. Using
this program one should be able to compute the critical temperature.
The main difficulty is that now one has to use a more complex set of
renormalization prescriptions to determine $a_s(b_0,b_s)$ and
$a_0(b_0,b_s)$ separately.

\begin{figure}[ht]
\centering
\begin{subfigure}{.5\textwidth}
  \centering
    \includegraphics[width=.8\linewidth]{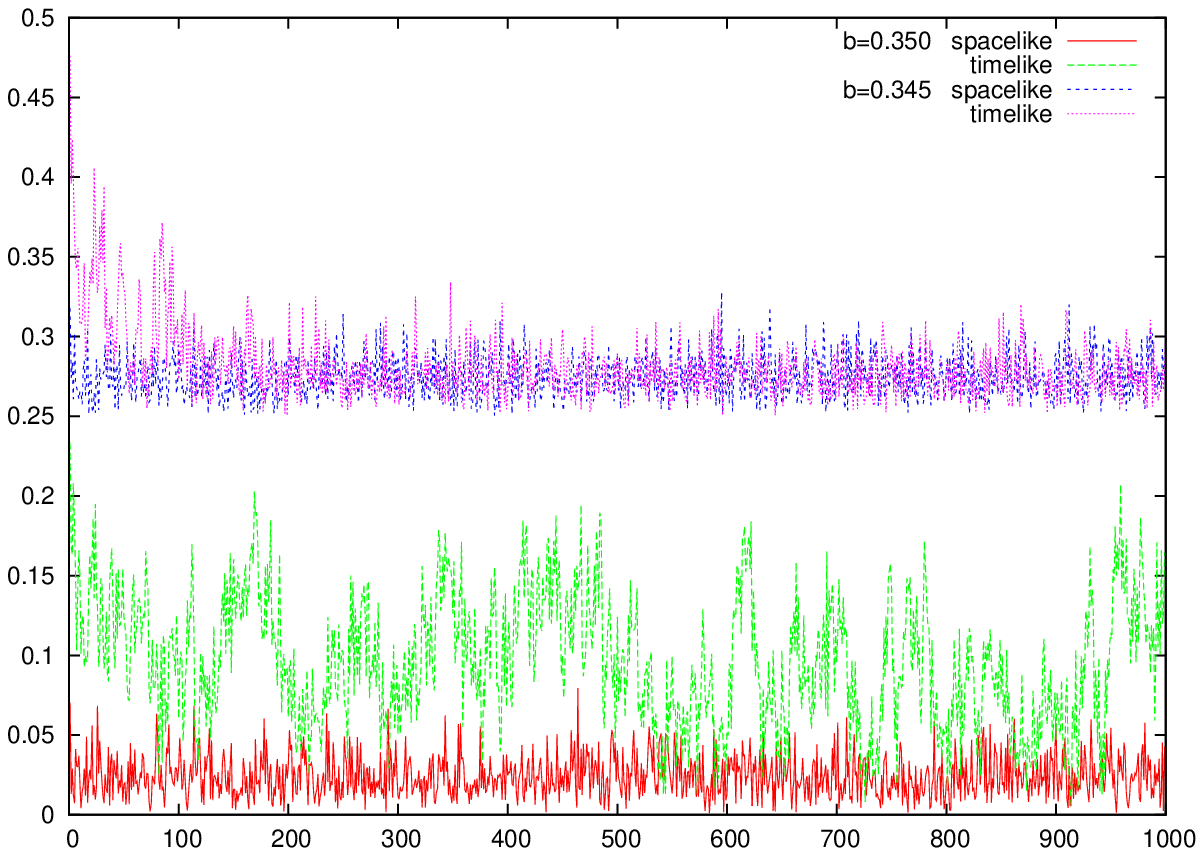}
      \caption{Fig. 1}
        \label{fig:fig1}
	\end{subfigure}%
	\begin{subfigure}{.5\textwidth}
	  \centering
	    \includegraphics[width=\linewidth]{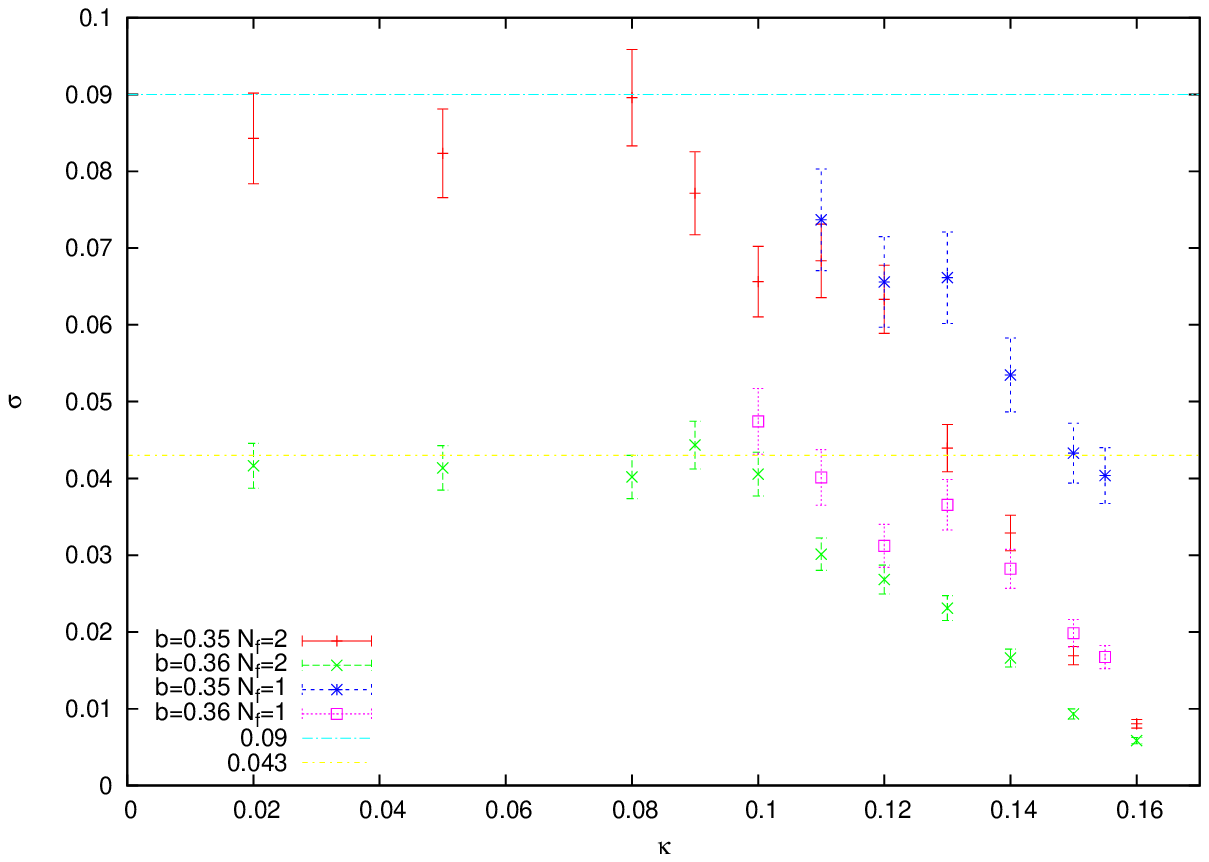}
	      \caption{Fig. 2}
	        \label{fig:fig22}
		\end{subfigure}
		\label{fig:test}
		\end{figure}


\section{Adding $N_f$ flavours of quarks in the adjoint}
The validity of reduction does not restrict to pure gauge theory.
Indeed, on the contrary, it was proposed~\cite{KUY} that 
in theories with quarks in the adjoint representation the quark loops 
might help to restore the symmetry breaking even when using strictly
periodic boundary conditions $k=0$. The arguments are based on
calculations done on  $S_1\times R^3$. The length of the periodic
direction $l_0$ becomes irrelevant in the large N limit, a phenomenon
named {\em volume independence}. Indeed, for small $l_0$ one can
reliably compute the effective potential for the Polyakov loops.
Gauge field contributions induce a breaking of $Z(N)$ symmetry, 
while quark loops tend to restore it. This is strictly so for massless 
quarks. However, it has been claimed that symmetry restoration also 
applies in certain region with cut-off scale quark masses~\cite{AHUY}. 
If this is so, quarks disappear  from the low energy dynamics and one 
has simply an implementation of reduced pure gauge theory. 

However, the gauge theory with dynamical quarks in the adjoint
representation is interesting in its own right and has attracted
considerable interest lately. Indeed, it is one of the 
candidates for a walking technicolor theory~\cite{DS}. The gauge theory 
with $N_f=2$ flavours of quarks in the adjoint is expected to lie
within the conformal window. Several studies have been carried for the
SU(2) gauge theory on the lattice consistent with this fact 
(See for example Ref.~\cite{DDB}). The same is expected to hold for
the large N counterpart. Furthermore, the large N gauge theory has
many other attractive features. For example, via orientifold planar 
equivalence, the theory matches an alternative large N limit of QCD
(two-index antisymmetric quark model).  Furthermore,  the case with 
a single Weyl fermion is supersymmetric. 

Studying  large N theory with dynamical fermions on the lattice
poses a formidable computational challenge. It is quite clear that 
reduction can be crucial to obtain results at an affordable
cost. For that reason, several authors have addressed the problem of 
simulating the corresponding reduced model for $N_f=1,2$, several
values of N and different versions of lattice fermions. One or two
flavours of overlap fermions have been studied by Hietanen and Narayanan~\cite{HN}.  
Other authors~\cite{BS}-\cite{BKS} use Wilson fermions. In all cases 
they show evidence that $Z(N)^4$ symmetry is preserved for a certain
range of quark masses and values of N. 

In the last few months we have addressed the problem from the
perspective of the twisted reduced model. As a matter of fact, the
fermionic part of the action is just that of a single site Wilson 
fermion coupled to the single site gauge field. The pure glue part is
given by $S_{\mathrm{TEK}}$ as before. Several values of the  integer 
flux parameter $k$ were used, including $k=0$ corresponding to the
untwisted case studied  in Ref.~\cite{BKS}. Our study is still in
progress, but a number of interesting conclusions have been obtained
already. As expected, for appropriate ranges of $k$, there is no sign
of symmetry breaking for arbitrary values of the quark mass and a range
of $N$ values going much beyond the one studied by previous authors. 
Furthermore, for large masses, the physical results obtained from
Creutz ratios  match with those obtained for the pure gauge theory. 
This confirms that, once twist is adopted, quark loops are an unnecessary 
ingredient in selecting the correct symmetry invariant vacuum. 

A strong conclusion which we obtained from our work~\cite{AGAMOADJ}
is that the magnitude of the N-dependence of the results depends
crucially on the value of $k$. For example, the plaquette expectation
value for the untwisted $k=0$ case is strongly  dependent on N with a pattern which is
sometimes non-uniform. This casts serious doubts on the usefulness of
the $k=0$ results to explore the physics of the large N model. On the
contrary, the  results for large enough $k$ values have quite small
N-dependence. In the appropriate range of masses it seems, however,
that the large N value of the plaquette is independent of $k$.

We have been able to obtain results up to $N=289$. This corresponds to
an effective box of length $L=17$. This value is large enough to allow
the study of Wilson loops up to   $8\times8$ size. A rough  estimate  of
the string tension can be extracted using similar techniques as for the  
the pure gauge case. In addition to smeared Wilson loops, we also
studied the lowest eigenvalue $\lambda$ of the square of the hermitian 
Wilson Dirac operator  $(\gamma_5 D_W)^2$. The eigenvalue decreases
with the hopping parameter $\kappa$, and seems to vanish at a critical
value $\kappa_c$.  Indeed, the data is well described by a formula 
\begin{equation}
\label{fit_formula}
\propto (\frac{1}{\kappa}- \frac{1}{\kappa_c})^\rho 
\end{equation}
The best fit parameters are summarized in the Table, for two values of
$b$ and $N_f=1,2$. We expect that the critical value signals the
vanishing of quark masses.

In Fig. 2 we display the value of the string tension for $b=0.35, 0.36$ 
and $N_f=1,2$. For small $\kappa$ (large bare masses) the
string tension matches with the one corresponding to the pure gauge
theory. The horizontal lines denote the value obtained from TEK and 
$N=841$. At around $\kappa\sim 1$ the string tension starts to depend
strongly on $\kappa$. From the plot one sees an approximate linear
behaviour. For $N_f=2$ the string tension seems to vanish around
$\kappa_c$. A fit of the form Eq.~\ref{fit_formula} gives good
$\chi^2$ and $\rho$ close to 1. Indeed setting $\rho=2/(1+\gamma^*)$
the fit gives the values of $\gamma^*$ plotted in the Table. 
The behaviour for $N_f=1$, although also consistent with linearity, does
not appear to vanish at $\kappa_c$.

\begin{table}
\begin{tabular}{||l||l|l|l|l||l|l|l|} \hline
 $b$ & $N_f$ &  $\kappa_c$ & $\rho$ & $\gamma^*$ & $N_f$ &   $\kappa_c$ & $\rho$ \\ \hline \hline
 0.035 & 2 & 0.1694(7) & 1.28(3) & 1.00(5) & 1  & 0.173(1) & 1.25(3)\\ \hline
 0.036 & 2 & 0.168(1) & 1.33(3) & 1.3(2) & 1 &  0.171(1) & 1.28(3)  \\ \hline
\end{tabular}
\end{table}

\section{Conclusions and outlook}
In the previous sections we have studied various large N theories using 
the reduction idea. We have shown that computations are possible and,
whenever the comparison could be made, match the results obtained by
more conventional techniques. The numerical coincidence of the string
tension implies that whatever mechanism drives Confinement it must be
well described by our matrix model. As is well-known,  the microscopic
mechanism driving Confinement for pure Yang-Mills theory  is not yet 
fully understood. This is so, despite the fact that we know since the
early 70's that Confinement is a possible phase of gauge theories.
To understand the meaning of microscopic mechanism, it is good to draw
a parallelism with  {\em superconductivity}. While condensation of a
charged field is always at work, the BCS theory provides the microscopic
mechanism that drives the condensation. For high temperature 
superconductors this theory is lacking.

In principle, the microscopic mechanism is specific of each theory and 
each dimensionality. We reject the common practice of exporting from
one theory to other. This, however, does not apply to changes in N.  
 As we saw, the string tension evolves mildly with N. Thus, there must
 also be a smooth continuity in the actual mechanism responsible for
 it, and a simple connection with the matrix model.

A.G-A is supported by grants  FPA2009-08785, FPA2009-09017,
CSD2007-00042, HEPHACOS S2009/ESP-1473 and ITN PITN-GA-2009-238353. 
M. O. is supported by the Japanese MEXT grant No 23540310.
Calculations were done on SR16000 computer at KEK supported by 
the Large Scale Simulation Prog. No.12-01 (FY2011-12).




\end{document}